\documentclass{llncs}
\def\ifundefined#1{\expandafter\ifx\csname#1\endcsname\relax}
\ifundefined{preambleloaded}% precompiled preamble a la
\usepackage{xcolor}
\usepackage{amsmath}
\usepackage{amssymb}
\usepackage{mathtools}
\usepackage{float}
\usepackage{tikz}
\usepackage{pgfplots}
\usetikzlibrary{backgrounds,fit,decorations.pathreplacing,snakes,arrows,shapes,automata}
\usetikzlibrary{patterns}
\usepackage{ifthen}
\usepackage{ifpdf}
\usepackage{easpc}
\usepackage{qex}
\usepackage{lskframe}
\usepackage{algorithm}
\usepackage[noend]{algpseudocode}

\ifpdf
\relax
\else
\usepackage{pstricks,pst-node}
\fi

\usepackage{chngcntr}

\counterwithin*{equation}{section}
\counterwithin*{equation}{subsection}

\title{From Stateless to Stateful Priorities: \\Technical Report}
\author{Christian Herrera}
\institute{fortiss GmbH, Munich Germany}
\makeatletter
\newcommand{\todo@box}[1]{\fcolorbox{green!50!black}{green!50!white}{#1}}
\newcommand{\todo}[1]{%
  \begin{center}
  \normalfont%
  \todo@box{\parbox{0.9\textwidth}{\color{black}{\bf TODO}: #1}}
  \end{center}}
\makeatother

\newcommand{\ofsane}[1]{\ifthenelse{\equal{#1}{(}}{\errmessage{Hups, '(' as of-macro parameter?}}{}}
\newcommand{\cnf}{\copyright}

\newcommand{\clk}{x}

\newcommand{\vars}{V}
\newcommand{\var}{v}
\newcommand{\expr}[1][int]{\psi_\mathit{#1}}
\newcommand{\Xconstrs}[2][]{\Phi_{#1}(#2)}
\newcommand{\Yconstrs}[2][]{\Psi_{#1}(#2)}
\newcommand{\constr}[1][]{\varphi\ifthenelse{\equal{#1}{}}{}{_#1}}

\newcommand{\vconstrs}[1][\vars]{\Xconstrs{#1}}
\newcommand{\vexpress}[1][\vars]{\Yconstrs{#1}}

\newcommand{\val}{\nu}
\newcommand{\assig}{\mu}
\newcommand{\dbs}[1]{[\![\hspace{0.04cm}#1\hspace{0.08cm}]\!]}

\newcommand{\locs}{L}

\newcommand{\loc}{\ell}

\newcommand{\iloc}[2][]{\loc_{\mathit{ini{#2}\ifthenelse{\equal{#1}{}}{}{#1}}}}
\newcommand{\nstloc}[2][]{\loc_{\mathit{nst{#2}\ifthenelse{\equal{#1}{}}{}{#1}}}}

\newcommand{\acts}{B}

\newcommand{\act}{\alpha}
\newcommand{\locinv}{I}
\newcommand{\edges}{E}
\newcommand{\edge}{e}

\newcommand{\transprog}{T}

\newcommand{\progpair}{\langle\locinv,\transprog\rangle}
\newcommand{\ta}{\mathcal{A}}

\newcommand{\theory}{\mathcal{C}} 

\newcommand{\datuple}[1][]{(\locs_{#1},\acts_{#1},\vars_{#1},\edges_{#1},\iloc{})}

\newcommand{\us}{\mathcal{U}(\vars)}

\newcommand{\rvec}{\vec{u}}
\newcommand{\lvec}{\vec{\loc}}
\newcommand{\true}{\mathit{true}}

\newcommand{\edgetuple}[1][]{(\loc_{#1},\act_{#1},\constr_{#1},\rvec_{#1},\loc'_{#1})}

\newcommand{\locsof}[1]{\ofsane{#1}\locs(#1)}
\newcommand{\edgesof}[1]{\ofsane{#1}\edges(#1)}

\newcommand{\actsof}[1]{\ofsane{#1}\acts(#1)}

\newcommand{\lts}[1]{\ofsane{#1}\mathcal{T}(#1)}

\newcommand{\states}[1]{\ofsane{#1}\mathit{Sts}(#1)} 
 
\newcommand{\tsconf}{s}

\newcommand{\iconfs}[1][]{\mathcal{C}_\mathit{ini\ifthenelse{\equal{#1}{}}{}{,#1}}}
\newcommand{\istates}[1][]{\sta_\mathit{ini\ifthenelse{\equal{#1}{}}{}{,#1}}}

\newcommand{\trans}[1]{\xrightarrow{#1}}
\newcommand{\texttrans}[1]{\mathrel{\smash[t]{\trans{#1}}}}
\newcommand{\locof}[1]{\ofsane{#1}\loc_{#1}}
\newcommand{\valof}[1]{\ofsane{#1}\val_{#1}}
\newcommand{\confln}[1]{\langle\lvec_{#1}, \valof{#1}\rangle}
\newcommand{\preerror}{(\mathit{loc},\mathit{var},\mathit{act},\mathit{stp})}
\newcommand{\varslocs}[3][]{\vars^{\ta}({\dbs{\dot{\nta}}^{#1}_{#2}},#3)}
\newcommand{\varsvars}[3][]{\vars^{\var}({\dbs{\dot{\nta}}^{#1}_{#2}},#3)}
\newcommand{\varsacts}[3][]{\vars^{\act}({\dbs{\dot{\nta}}^{#1}_{#2},#3})}

\newcommand{\comp}{\sigma}
\newcommand\pto{\mathrel{\ooalign{\hfil$\mapstochar$\hfil\cr$\to$\cr}}}

\newcommand{\paths}[1]{\mathit{Paths}(#1)}
\newcommand{\nat}{\mathbb{N}}
\newcommand{\natplus}{\nat^{>0}}

\newcommand{\nta}{\mathcal{N}}

\newcommand{\sta}{\mathcal{S}}

\newcommand{\repifygrd}{\Gamma}

\newcommand{\alg}{\mathcal{K}}
\newcommand{\algprio}{\alg^{\rho}}
\newcommand{\posvar}[1]{\mathit{p}_{#1}}
\newcommand{\reachconf}[1]{\mathit{Reach}_{#1}}

\newcommand{\BF}{\beta}

\newcommand{\SF}{\mathit{SF}}

\newcommand{\epfq}[1]{\mathop{\existsOooginool\Diamond} #1}

\newcommand{\prio}{(\act_1,\act_2)}

\def\preambleloaded{Precompiled preamble loaded.}
\else\fi

\begin{document}

\maketitle

\begin{abstract}
We present the notion of \emph{stateful priorities} for
impo\-sing precise restrictions on system actions, in order to meet safety cons\-traints. 
By using stateful priorities we are able to 
exclusively restrict erroneous system behavior as specified by the constraint, 
whereas safe system behavior remains unrestricted. 
Given a system modeled as a network of discrete automata and an error constraint, 
we present algorithms which use those inputs to synthesize stateful priorities. 
We present as well a network transformation which uses synthesized prio\-rities 
for blo\-cking all system actions leading to the input error.
Our experiments with three real-world examples demonstrate the applicabi\-lity of our approach\-.
\end{abstract}

\section{Introduction}
\label{sec:intro}

Using \emph{stateless priorities}~\cite{Bornot,Cleaveland2} is a common practice 
for imposing global res\-trictions on system actions, and thereby influencing the behavior of distributed systems
in order to meet given constraints. This practice is particularly useful in domains like mutual exclusion~\cite{Bornot}, 
fault-repair~\cite{Cheng} and conflict reso\-lution~\cite{Sifakis}. 
%Distributed systems with priorities have been modeled in 
%process algebra~\cite{Cleaveland}, timed automata~\cite{Herrera5} and hybrid automata~\cite{Bornot3}, among others.

For distributed systems intended to meet specific constraints, e.g.\ to avoid a particular error state denoted by
two or more components entering their criti\-cal section at the same time, 
using stateless priorities often imposes strong restrictions with two significant consequences:
(1) disabling safe system behavior and, 
(2) inducing unnecessary verification overhead. 
For instance, stateless priorities disable safe behavior in distributed systems, 
if actions from safe states are res\-tricted although they do not lead to the error state. 
Note that this often restricts as well reachability of safe states.
The unnecessary overhead is caused by applying stateless priorities on actions from safe 
states that lead to safe states. 
Unne\-cessary computations are performed in order to determine, (a) all enabled actions (if any) at a parti\-cular safe state and, 
(b) the order for executing ena\-bled actions while respecting the underlying priorities. 
Models using priorities in verification tools like \emph{Uppaal}~\cite{Uppaal} and \emph{BIP}~\cite{Basu2} suffer from these consequences, 
since the implementation of priorities in those tools can be considered stateless as they impose global restrictions on system actions.
Note that the application of priorities in BIP can be conditioned, however this is still inadequate for avoiding the mentioned consequences.
%For an analysis of the contributions to memory consumption and verification time of priorities in Uppaal models,
%we refer the interested reader to~\cite{Herrera6}.

The unnecessary overhead and disablement of safe system behavior can be avoided by using \emph{stateful priorities}.
Intuitively, a stateful priority is a pair consisting of a state which is one transition step away from reaching the error,
and a priority which from that state restricts an action that leads to the error.  

We present a set of algorithms which use three inputs for synthesi\-zing stateful priorities.
The first input is a distributed system mode\-led as a network of discrete automata. Our mode\-ling language 
is rich enough to model real-world examples. The second input is an error constraint expressed as a conjunction of 
automata locations. Interestingly, conjunctions of locations are sufficient for expressing
error states in each of our real-world examples, even those errors which are naturally expressed with data variables.
The third input is a bound on the number of verification steps. 
We present as well a network transformation that make guards of edges more restrictive, 
by adding integer \emph{positional} variables that make use of synthesized priorities.
Intuitively, positional variables rule out states from which action transitions lead to the error.
The result is a network where erroneous system behavior is precisely restricted, whereas safe system behavior remains unrestricted.
%Moreover, our transformation does not introduce new \emph{deadlocks}, i.e.\ states from which no successor states 
%are induced by transitions. 
We provide our approach as a source-to-source transformation which yields models that 
can be easily translated into Uppaal and BIP models, and further verification techniques can be directly applied.

In summary, our contributions are: 
(1) the notion of stateful priorities which allows to precisely res\-trict erroneous system behavior, while safe system behavior remains unrestricted, 
(2) a set of algorithms for synthesi\-zing stateful priorities, and a network transformation which uses those priorities for
restricting erroneous behavior and, 
(3) an automatic source-to-source transformation of models.
 
This paper is organized as follows. Section~\ref{prelim} provides basic definitions.
Section~\ref{running} introduces an example.
Section~\ref{encoding} provides encodings for networks of discrete automata and error cons\-traints.
Section~\ref{transfo} provides algorithms for synthesizing stateful priorities, and introduces a network transformation
for using those priorities.
Section~\ref{correctness} shows the correctness of our approach.
Section~\ref{exp} presents our experiments. 
Section~\ref{related} presents related work and conclusions.

\section{Preliminaries}
\label{prelim}

%We use the following definitions. 
%\medskip
%
Let $\vars^r$ be a set of \emph{real varia\-bles}. 
Let $\vexpress[\vars{^r}]$ be a set of real \emph{expressions} defined by the usual syntax using variables in $\vars^r$, and the function symbols $+,-,\dots$
Let $\vconstrs[\vars{^r}]$ be a set of real \emph{constraints} defined by the usual syntax using variables in $\vars^r$, real expressions, 
the predicate symbols: $<,\le,=,\ge,>$, and the logical connectives: $\land,\neg,\lor$.
We assume the canonical satis\-faction relation ``$\models$'' between 
\emph{valua\-tions} $\val : \vars^r \to \mathbb{R}$ and real constraints.
Let $\vars^b$ and $\vars^{int}$ be sets of \emph{boolean} and \emph{integer} varia\-bles, respectively.  
Sets of expressions and constraints for boolean and integer variables, as well as their satisfaction relations are defined similarly. 
Let $\vars= \vars^r\cup\vars^b\cup\vars^{int}$. 
%Note that variables in $\vars$ can be shared.
An \emph{update vector} $\rvec\in\us$ is a finite and possibly empty sequence of \emph{assignments}, $\var := \expr[]$,
where $\var \in \vars$ and $\expr[]\in\vexpress[\vars]$. 
%Let $\us$ be a set of update vectors.
%Let $\bcacts$ be a set of actions.
%
A (discrete) automaton $\ta$ is a tuple $\datuple$ which consists of
a finite set of \emph{locations} $\locs$, where $\iloc{}\in\locs$ is the initial location, 
a finite set $\acts$ of actions, 
%comprising the \emph{internal action} $\tau$, $\tau\notin\bcacts$,
%
%
and a set of \emph{edges}
$\edges \subseteq 
\locs \times \acts \times \vconstrs \times \us \times \locs$.
An edge $\edge = \edgetuple \in \edges$ from location $\loc$ to $\loc'$ involves an
action $\act \in \acts$, a \emph{guard} $\constr \in \vconstrs$, and an
update vector $\rvec \in \us$.
%
%
%and $\var := \expr[boolean]$ where $\var \in \vars^b$ and $\expr[boolean]\in\mathbb{B}$.
%
We write $\iloc{}(\ta)$, $\acts(\ta)$, $\edges(\ta)$, etc.\ to denote the
the initial location, the set of actions, the set of edges, etc.\ of $\ta$.
%We use $\chan(\act)$ 
%and $\chan(\acts)$ 
%to denote the action of a single action $\act$
%, and the actions 
%used from $\acts$. 
%Let $\varsigma: \bcacts\to\mathbb{B}$ be a function that 
%assigns a boolean value to each action in $\bcacts$. 
%Let $\varrho: \bcacts\to\bcacts$ be a renaming function that takes a action $\chan(\act)$ as input, and outputs
%$\chan(\act)^\ta$ if $\varsigma(\chan(\act))$ equals true, and $\chan(\act)$ otherwise. 
%Let $\omega: \vars\to\mathbb{B}$ be a function that assigns a boolean value to each variable in $\vars$.

A \emph{network $\nta$ (of automata)} consists of
a finite set $\{\ta_1, \dots, \ta_N\}$ of automata with pairwise disjoint sets of 
locations.
%$\bcacts^{\mathit{b}} \subseteq \bigcup_{i=1}^N\actsof{\ta_i}$ of
%\emph{broadcast actions} where broadcast actions model one-to-many synchronization.
We write $\ta \in \nta$ if and only if $\ta \in \{ \ta_1, \dots, \ta_N \}$.
The set of states
$\states{\nta}$ consists of pairs of \emph{location vectors}
$\langle \loc_1, \dots, \loc_N \rangle$
from $\times_{i=1}^N \locsof{\ta_i}$,
and valuations of $\bigcup_{i=1}^N\vars(\ta_i)$.
We use $\locof{\tsconf,i}$, $1 \leq i \leq N$, to denote the location which
automaton $\ta_i$ assumes in state $\tsconf = \confln{s}$.
%
%Let $\rho: \bcacts\cup\{\tau\}\to\nat$ be a function that 
%assigns a \emph{priority} to $\tau$ and to each action in $\bcacts$. 
%
%
The set of initial states $\istates$ consists of $\istates=\{\langle\lvec_{ini},\{\val_{ini}\}\rangle\} \cap \states{\nta}$,
  where $\lvec_{ini}=\langle \loc_{ini,1}, \dots, \loc_{ini,N} \rangle$ and
  $\val_{ini}$ assigns (user predefined) initial va\-lues to each $\var\in\vars(\ta_i)$. 
The concrete semantics of the network $\nta$ is given by the transition system
$\lts{\nta}=(\states{\nta},\actsof{\ta_1}\cup\cdots\cup\actsof{\ta_N},
\{\texttrans{\act} \mid \act \in \actsof{\ta_1}\cup\cdots\cup\actsof{\ta_N}\}, \istates)$.
Between two states $\tsconf,\tsconf'\in \states{\nta}$ there exists:
\begin{itemize}
\item \emph{action transition} (or single transition)
$\confln{s} \texttrans{\act} \langle\lvec_{\tsconf'}, \val_{\tsconf'}\rangle$, 
if for some $1 \leq i\leq N$, and an edge 
$(\loc_i,\act,\constr_i,\rvec_i, \loc'_i) \in \edgesof{\ta_i}$, with $\act\in\acts(\ta_i)$, in the $i$th automaton
such that:
$\lvec_{s'}=\lvec_s[\loc_{s,i}:=\loc'_i]$, i.e.\ location updated;  
$\val_{\tsconf}\models\constr_i$, i.e.\ guard satisfied; 
and $\val_{\tsconf'}:=\val_{\tsconf}[\rvec_i]$, i.e.\ variables updated by update vector $\rvec_i$,

\item 
\emph{broadcast transition} (or synchronization transition)
$\confln{s} \texttrans{\act} \langle\lvec_{\tsconf'}, \val_{\tsconf'}\rangle$,
if there exist indices, 
$1 \leq i_1, \dots, i_k \leq N$, with $k>1$,
and if there exists an action  
$\act \in \acts(\ta_{i_1})\cap\cdots\cap\acts(\ta_{i_k})$ such that there exist edges 
$(\loc_{i_j},\act,\constr_{i_j},\rvec_{i_j}, \loc'_{i_j}) \in \edgesof{\ta_{i_j}}$, for all $i_j\in\{i_1, \dots, i_k\}$,
such that: $\lvec_{\tsconf'}=[\loc_{\tsconf,i_1}:=\loc'_{\tsconf,i_1}]\cdots[\loc_{\tsconf,i_k}:=\loc'_{\tsconf,i_k}]$,
$\val_{\tsconf}\models\constr_{i_1}\land\dots\land\constr_{i_k}$ and $\val_{\tsconf'}:=\val_{\tsconf}[\rvec_{i_1}]\dots[\rvec_{i_k}]$.

\end{itemize}
A finite or infinite sequence 
$
\comp = \tsconf_0\trans{\act_1} \tsconf_1 \trans{\act_2} \tsconf_2\cdots
$ 
of states is called \emph{transition sequence}, with $\tsconf_0\in \istates$,
of $\nta$.
Sequence $\comp$ is called \emph{computation path} of $\nta$ if and only
if it is finite and $\tsconf_0 \in \istates$.
$\paths{\nta}$ denotes the set of all computation paths of $\nta$.
A state $\tsconf$ is called \emph{reachable}
(in $\lts{\nta}$) if and only if there exists 
$\comp \in \paths{\nta}$ such that $\tsconf$ occurs in $\comp$. 
The set $\reachconf{\nta}$ (or state space) contains 
all reachable states of $\nta$\-.
A reachable state $\dot\tsconf$ is called \emph{deadlock} if and only if 
no successor state $\dot\tsconf'$ is induced by any kind of transition from $\dot\tsconf$.
We write $\locs(\nta)$, $\vars(\nta)$  etc.\ to denote the set of locations, variables, etc.\ of $\nta$.

The set of \emph{basic formulae} over $\nta$ is given by the grammar
$\BF ::= \ta.\loc \mid \neg\ta.\loc$ 
where $\ta\in\nta$ and $\loc \in \locs(\ta)$. 
%We use $\BF_{\ta}$ and $\BF_{\loc}$
%to refer to the automaton and the location ocurring in $\BF$, respectively.
%
Basic formula $\BF$ is satisfied by state $\tsconf \in \states{\nta}$,
if and only if
$\locof{\tsconf,i} = \ell$, or $\locof{\tsconf,i} \neq \ell$, with $1 \leq i \leq N$.
A \emph{reachability query} over $\nta$ is $\epfq{\SF}$ where $\SF$ is
a \emph{state formula} over $\nta$, i.e.\ 
any conjunction of basic formulae.
%
%We use $\BFof(\CF)$ to denote the set
%of basic formulae in $\CF$.
%
$\nta$ satisfies $\epfq{\SF}$,
 denoted by $\nta \models \epfq{\SF}$, 
if and only if there is a state $\tsconf$ reachable in $\lts{\nta}$ s.t.\ 
$\tsconf \models \SF$.

\section{Running Example}
\label{running}

\begin{figure}[t]
\centering
\begin{tikzpicture}[>=latex',join=bevel,]
\tikzstyle{every state}=[draw=blue!50,very thick,fill=blue!20]%
\tikzstyle{state2}=[draw=white,fill=white]%

\node (A0) at (60bp,110bp) [state2] {};
\node (A1) at (37bp,87bp) [state] {\scriptsize $1$};
\node (A2) at (100bp,87bp) [state] {\scriptsize $2$};
\node (A3) at (160bp,87bp) [state] {\scriptsize $3$};
\node (A4) at (80bp,57bp) [state] {\scriptsize $4$};
\node (A5) at (160bp,57bp) [state] {\scriptsize $5$};
\node (F2) at (37bp, 108bp)  {$\ta_0$:};

\draw [->] (A0) to[]  node[above] {}  node[below,xshift=0.60cm, yshift=0.35cm] {\scriptsize $\clk:=1$}(A1); 
\draw [->] (A1) to[]  node[above] {\scriptsize $a$}  node[below] {\scriptsize $\clk:=x+1$}(A2); 
\draw [->] (A2) to[]  node[above] {\scriptsize $a$}  node[below] {\scriptsize $\clk:=x+2$}(A3);
\draw [->] (A1) to[out=270]  node[above, xshift=-0.21cm, yshift=-0.08cm] {\scriptsize $e$}  node[below] {}(A4);
\draw [->] (A4) to[]  node[above] {\scriptsize $a$}  node[below] {\scriptsize $\clk:=x-1$}(A5);
\draw [->] (A5) to[bend left, looseness=1.7]  node[above] {\scriptsize $b$}  node[below] {\scriptsize $\clk:=x+1$}(A4);

\node (B0) at (240bp,110bp) [state2] {};
\node (B1) at (217bp,87bp) [state] {\scriptsize $1$};
\node (B2) at (280bp,87bp) [state] {\scriptsize $2$};
\node (B3) at (340bp,87bp) [state] {\scriptsize $3$};
\node (B4) at (260bp,57bp) [state] {\scriptsize $4$};
\node (B5) at (340bp,57bp) [state] {\scriptsize $5$};
\node (G2) at (217bp, 108bp)  {$\ta_1$:};

\draw [->] (B0) to[]  node[above] {}  node[below,xshift=0.60cm, yshift=0.35cm] {\scriptsize $\clk:=1$}(B1); 
\draw [->] (B1) to[]  node[above] {\scriptsize $d$}  node[below] {}(B2); 
\draw [->] (B2) to[]  node[above] {\scriptsize $d$}  node[below] {}(B3);
\draw [->] (B1) to[out=270]  node[above, xshift=-0.21cm, yshift=-0.08cm] {\scriptsize $e$}  node[below] {}(B4);
\draw [->] (B4) to[]  node[above] {\scriptsize $c$}  node[below] {\scriptsize $\clk:=x-1$}(B5);
\draw [->] (B5) to[bend left, looseness=1.7]  node[above] {\scriptsize $d$}  node[below] {\scriptsize $\clk:=x+1$}(B4);
\draw [->] (B3) to[loop, looseness=1, loop, distance=1cm]  node[above] {\scriptsize $d$}  node[below] {}(B3);
\end{tikzpicture}

\vspace*{-1em}
\caption{Network $\nta_1$ of discrete automata with global integer variable $\clk$.}
\label{fig1}
\end{figure}
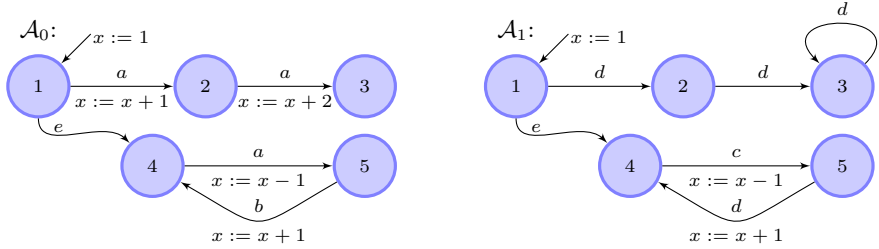

In this section we introduce an example of a network of discrete automata, 
and we give an intuition of our approach which uses two important notions, \emph{preErrors} and \emph{priorities}.
Intuitively, a preError is a state exactly one transition step away from reaching a given error state. A
priority is a pair of actions denoting a preference for executing transitions labeled with those actions. 

\begin{definition}[PreError] 
Let $\nta$ be a network. 
Given an error state $\tsconf$, and a computation path reaching $\tsconf$, i.e.\
$\tsconf_0\trans{\act_1}\cdots\trans{\act_{n-1}}\tsconf_{n-1}\trans{\act_n}\tsconf_n=\tsconf\in\paths{\nta}$,
then we call the state $\tsconf_{n-1}$ \emph{preError}. 
\qex
\label{preerror}
\end{definition}

\begin{definition}[Priority] 
A \emph{priority} $\rho$ is a pair $\prio$ of actions of network $\nta$, denoting that
whenever two sets of edges $\edges_{\act_1} = \{\edge_1,\dots,\edge_n\}$ such that $\act_1$ is the action for each $\edge\in\edges_{\act_1}$, and 
$\edges_{\act_2} = \{\dot\edge_1,\dots,\dot\edge_m\}$ such that $\act_2$ is the action for each $\dot\edge\in\edges_{\act_2}$, are enabled,
then all edges in $\edges_{\act_2}$ must be taken before any edge in $\edges_{\act_1}$ is taken. 
From a priority $\rho=\prio$ we call $\act_1$ \emph{blockee} and use $\rho_{\mathit{be}}$ to refer to it, 
and $\act_2$ \emph{blocker} and use $\rho_{\mathit{br}}$ to refer to it.
\qex
\label{priority}
\end{definition}

\begin{example}
Figure~\ref{fig1} shows network $\nta_1$ consisting of $\ta_0$ and $\ta_1$.
Assume that any state where the value of $\clk$ is negative denotes an error for $\nta_1$.
Variable $\clk$ becomes nega\-tive whenever $\ta_0$ and $\ta_1$ are located at the same time at their locations $5$,
thus, our error formula is $\phi:=\ta_0.5 \land \ta_1.5$.
For reaching $\phi$ in $\nta_1$ we show computation paths
\small$\comp_1=\langle(\ta_0.1,\ta_1.1),\valof{}(\clk)=1\rangle\trans{e}
\langle(\ta_0.4,\ta_1.4),\valof{}(\clk)=1\rangle\trans{a}\langle(\ta_0.5,\ta_1.4),\valof{}(\clk)=0\rangle\trans{c}
\langle(\ta_0.5,\ta_1.5),\valof{}(\clk)=-1\rangle$\normalsize, and 
\small$\comp_2=\langle(\ta_0.1,\ta_1.1),\valof{}(\clk)=1\rangle\trans{e}
\langle(\ta_0.4,\ta_1.4),\valof{}(\clk)=1\rangle\trans{c}\langle(\ta_0.4,\ta_1.5),\valof{}(\clk)=0\rangle\trans{a}
\langle(\ta_0.5,\ta_1.5),\valof{}(\clk)=-1\rangle$\normalsize.

Our approach performs three main steps
in order to avoid reaching $\phi$ in all computation paths of $\nta_1$. 
We present the first two steps, the third is presented in Section~\ref{trafoNet}. 
Step 1 checks whether or not a state denoted by $\phi$ is reachable, 
if it is, then we collect all reachable preErrors. 
In our example we only have two preErrors, 
i.e.\ \small$\tsconf_1=\langle(\ta_0.5,\ta_1.4),\valof{}(\clk)=0\rangle\ $\normalsize  and
\small$\tsconf_2=\small\langle(\ta_0.4,\ta_1.5),\valof{}(\clk)=0\rangle$\normalsize.
Step 2 uses each reachable preError for synthesizing priorities (if any). 
Considering the preError $\tsconf_1$, 
only two action transitions are enabled from this state, 
i.e.\ $b$ and $c$. Performing the $c$-transition leads directly to the error state, therefore the transition with $b$ is preferred over $c$, 
and thereby the error state is avoided. This yields our first priority, $(c,b)$. 
Alternatively, from the preError $\tsconf_2$,
only two transitions are enabled, i.e.\ $a$ and $c$. Performing the $a$-transition leads directly to the error state, 
therefore the transition with $d$ is preferred over $a$, 
and thereby the error state is avoided. This yields our second priority, $(a,d)$. 
\label{example1}
\qex
\end{example}

Using those two priorities in Uppaal and BIP avoids reaching $\phi$. However, 
using $(a,d)$ in those tools restricts   
$\ta_0$ from reaching locations $2$ and $3$, whenever transitions with actions $a$ and $d$ are at the same time enabled. 
This restriction of behavior is too severe since these locations are safe.
Therefore, Section~\ref{trafoNet} presents a transformation that uses the information of preErrors and priorities
to restrict only transitions reaching $\phi$, while safe behavior remains unrestricted.

\section{Encoding of Networks}
\label{encoding}

We borrow the following definitions from~\cite{Sorea}. 
Let \emph{tt} (true) and \emph{ff} (false) be cons\-tants. 
Let $\theory$ be a boolean language associated to the set of variables $\vars\cup\{\mathit{tt},\mathit{ff}\}$ 
and closed under: $\land, \lor$ and $\neg$. 
%A set of constraints in $\lang$ is called a \emph{constraint theory} $\theory$ if it includes
%the constants \emph{tt} (true), \emph{ff} (false), and if it is closed under negation.
%A subset of $\theory$ of constraints with free variables in $\vars'\subseteq\vars$ is denoted by $\theory(\vars')$.
For a constraint $c\in\theory$ and an assignment $\assig$ for the free variables in $c$, 
the value of the predicate $\dbs{c}_\assig$ is called \emph{interpretation} of $c$ wrt.\ $\assig$. 
The interpretation $\dbs{\emph{tt}}_\assig$ ($\dbs{\emph{ff}}_\assig$) is assumed to hold for all (for no) $\assig$, 
and $\dbs{\neg c}_\assig$ holds if and only if $\dbs{c}_\assig$ does not hold.
Note that $\assig$ assigns respectively integer, real and boolean values to free integer, real and boolean variables in a given $c$, 
and preserves constant values, arithmetical and boolean operators.
A set of constraints $C\subseteq\theory$ is called \emph{satisfiable} if there exists an assignment $\assig$ such that $\dbs{c}_\assig$
holds for each $c\in C$; otherwise, $C$ is called \emph{unsatisfiable}. Moreover, a function $\theory$-$\mathit{sat}(C)$ is
called a $\theory$-satisfiability solver, it returns $\perp$ if the set of constraints are unsatisfiable, 
and a satisfying assignment $\assig_{\mathit{sat}}$ for $C$, otherwise. 
%We use $\dbs{\var}_{\assig_{\mathit{sat}}}$ to denote the 
%satisfying interpretation of each $\var\in\vars$ occurring in $C$.
%
%Assuming a solvable constraint theory $\theory$, variables in $\vars$ also denote \emph{state variables}, 
%$\vars'$ denotes a primed disjoint copy of $\vars$, and valuations over $\vars$ denote \emph{program states}. 
A $\theory$-program is a pair $\progpair$ where $\locinv$ denotes the set of initial states, 
and $\transprog$ the transition relation between states and their successor states.

%Before enconding networks we preprocess them in order to introduce actions with flexible semantics.
%This semantics transforms broadcast into action transitions, and thereby helps our reachability algorithms.
%In our tool CrEStO, users set boolean flags to define network actions with flexible semantics.
%This semantics (introduced by the following preprocessing) allow actions to participate in both, broadcast and action transitions.

\iffalse
\begin{definition}[Preprocessing of Networks]
Let $\nta = \{\ta_1, \dots, \ta_N\}$ be a network.
Let $\varsigma: \acts\to\mathbb{B}$ be a function that 
assigns a boolean value to each action in $\acts$. 
Let $\varrho: \acts\to\acts$ be a renaming function that takes an action $\act$ as input, and outputs
$\act^\ta$ if $\varsigma(\act)$ equals true, and $\act$ otherwise. 
Let $\nta' = \{ \alg(\ta_1), \dots,\alg(\ta_N)\}$,
where $\alg(\ta) = (\locs(\ta),\acts',\vars(\ta),\edges',\iloc{})$, 
$\acts' = \acts(\ta) \cup \{\varrho(\act) \mid \act\in\acts(\ta)\land \varsigma(\act)\}$, and 
$\edges'=\edges(\ta)\cup \{ (\loc, \varrho(\act), \constr, \rvec, \loc') \mid \edgetuple \in \edges(\ta) \land \varsigma(\act)\}$.
\qex
\end{definition}
\fi

The following encoding for networks of discrete automata 
allows us to use \emph{bounded model checking}~\cite{Clarke} as a technique for reachability analy\-sis.

\begin{definition}[Encoding of Networks] 
Let $\nta$ be a network.
%Let $\nta' = \{\alg(\ta_1),\dots,\alg(\ta_N)\}$.
Let $k\in\natplus$.
Let $\vars^{\act}_k:= \{\act^i\mid\act\in\acts(\nta), 0\le i\le k\}$ be a set of boolean variables.
Let $\vars^{\var}_k:= \{\var^i\mid \var\in\vars(\nta), 0\le i\le k\}$ be a set of real and boolean variables. 
Let $\vars^{\ta}_k = \{\ta^i\mid \ta\in\nta,0\le i\le k\}$ 
be a set of variables interpreted over $\locs(\nta)$.
Let $\bar{\vars} = \vars^{\act}_k\cup\vars^{\var}_k\cup\vars^{\ta}_k$. 
The encoding $\dbs{\dot{\nta}}_k$ of the $k$th unfolding of a 
$\theory$-program $\dot{\nta} = \progpair$ wrt.\ $\nta$ over $\bar{\vars}$, 
is given by the formula $\dbs{\dot{\nta}}_k:=\locinv\land\transprog$, where:
\vspace{-0.2cm}
\footnotesize
\begin{align}
{}&
\locinv:= \bigwedge\limits_{\ta\in\nta}\ta^0=\iloc{,\ta}\ \land  
	\bigwedge\limits_{\substack{\var\in\vars(\nta)}} \var^0 = \var(\rvec_{\mathit{ini}}), \label{init}\\
{}&
\transprog:= \bigwedge\limits_{i=0}^{k-1} \bigvee\limits_{\ta\in\nta}\Big(\hspace{-3.6cm}
\bigvee\limits_{\substack{\hspace{3.7cm}\edge=(\loc,\act,\constr,\langle\var_1:=\expr[1],\dots,\var_m:=\expr[m]\rangle,\loc')\in\edges(\ta)}}
\hspace{-3.5cm}(\ta^i = \loc \land \act^i \land \mathit{block}(\act,i) \land
\constr[][\var/\var^i\mid\var\in\vars(\nta)] \ \land \label{trans0}\\
{}&\bigwedge\limits_{\substack{\var_p\in\{\var_1,\dots,\var_m\}}} \hspace{-0.5cm}\var^{i+1}_p=\expr[p][\var/\var^i\mid\var\in\vars(\nta)] \ \land \hspace{-2.0cm}
\bigwedge\limits_{\substack{\hspace{2cm}\var\in\{\dot{\var}\in\vars(\ta)\mid \omega(\dot{\var})\}\setminus\{\var_1,\dots,\var_m\}}}\hspace{-2cm}
 \var^{i+1} = \var^i \land \ta^{i+1} = \loc') \ \lor\label{trans1}\\
{}&(\ta^{i+1} = \ta^{i} \land \bigwedge\limits_{\substack{\act\in\acts(\ta)}} \neg\act^i
\ \land \bigwedge\limits_{\substack{\var\in\{\dot{\var}\in\vars(\ta)\mid \omega(\dot{\var})\}}} \var^{i+1} = \var^{i}) 
\Big)\label{trans2}, 
\end{align}
\normalsize
and where: $\var(\rvec_{\mathit{ini}})$ denotes the initial value for each $\var\in\vars(\nta)$.
For an $\act\in\acts(\nta)$, $\mathit{block}(\act,i):=\bigwedge_{\bar\act\in\acts(\nta)\setminus\{\act\}} \neg\bar{\act}^i$,
with $\bar{\act}^i\in\vars^{\act}_k$, blocks transitions with actions different from $\act$. 
Function $\omega: \vars\to\mathbb{B}$ assigns true to each $\var\in\vars(\ta)$, with $\ta\in\nta$,
if $\var$ is exclusively updated in $\ta$, and false otherwise.
%We use $\varslocs{k}{i}$, $\varsvars{k}{i}$ and $\varsacts{k}{i}$
%to respectively denote the sets of variables from $\vars^{A}$, from $\vars^{\var}$ 
%and from $\vars^{\act}$ related to the $i$th unfolding, with $0\le i\le k$, occuring in constraints of $\dbs{\dot{\nta}}_k$. 
\qex
\label{encodingNet}
\end{definition}

From Definition~\ref{encodingNet}, constraint~$\ref{init}$ encodes the initial state. 
Constraints~$\ref{trans0}$ and $\ref{trans1}$ encode edges. For each edge $\edge$ of each automaton $\ta$, 
conjuncts of those cons\-traints encode in the following order:
origin location, action, actions to be blocked, guard, 
updates for variables updated by $\edge$, unchanged variables (because are not updated by $\edge$), and destination location.
Constraint~$\ref{trans2}$ encodes the fact of $\ta$ remaining idle while other automata perform transitions.

\begin{example}
For $k\in\{0,1\}$, we present an encoding $\dbs{\dot{\nta_1}}_k:=\locinv\land\transprog$ 
for the network $\nta_1$ of Figure~\ref{fig1}. 
We use the integer variables $\ta^k_0$ for automaton $\ta_0$, and $x^k$ for the integer variable $x$ in $\nta_1$.
For actions, we use the boolean variables: $a^k$, $b^k$, $c^k$, $d^k$ and $e^k$. We show only the first unfolding, 
and only for the edges in $\ta_0$. The remaining edges of $\ta_1$ can be easily encoded by following this example.
\scriptsize
\setcounter{equation}{0}
\begin{align}
{}&
\locinv:=  (\ta^0_0 = 1 \ \land \ \ta^0_1 =1 \ \land \ x^0=1 )\label{initEncoding}\\
{}&
 \transprog:= \Big((\ta^0_0 = 1 \ \land \ a^0 \ \land \bigwedge\limits_{\act\in\{b,c,d,e\}}\neg\act^0 \ \land \ x^1 = (x^0 + 1) \land \ \ta^1_0 = 2) \ \lor\label{c0} \\
{}&
(\ta^0_0 = 2 \ \land \ a^0 \ \land \ \bigwedge\limits_{\act\in\{b,c,d,e\}}\neg \act^0 \ \land \ x^1 = (x^0 + 2) \land \ \ta^1_0 = 3) \ \lor \label{c1}\\
{}&
(\ta^0_0 = 1 \ \land \ e^0 \ \land \ \bigwedge\limits_{\act\in\{a,b,c,d\}}\neg \act^0\ \land \ x^1 = x^0 \land \ \ta^1_0 = 4) \ \lor\label{c2} \\
{}&
(\ta^0_0 = 4 \ \land \ a^0 \ \land \ \bigwedge\limits_{\act\in\{b,c,d,e\}}\neg \act^0 \ \land \ x^1 = (x^0 - 1) \land \ \ta^1_0 = 5) \ \lor\label{c3} \\
{}&
(\ta^0_0 = 5 \ \land \ b^0 \ \land \ \bigwedge\limits_{\act\in\{a,c,d,e\}}\neg \act^0 \ \land \ x^1 = (x^0 + 1) \land \ \ta^1_0 = 4) \ \lor\label{c4} \\
{}&
(\ta^{1}_0 = \ta^{0}_0 \land \bigwedge\limits_{\substack{\act\in\{a,b\}}} \neg\act^0 \ \land \ x^1 = x^0)\Big) \ \land \ \cdots\label{c5}
\end{align}
\normalsize
Constraint~\ref{initEncoding} encodes the initial locations of both automata and the initial value of $x$.
Constraints~\ref{c0}-\ref{c5} encode the first unfolding for all edges of $\ta_0$. 
Note that from those constraints only one disjunct at a time can be satisfied, and this depends on the values of variables of the previous unfolding, 
in this example on the initial values encoded.
Considering the disjunct from constraint~\ref{c0}, if the conjunct $a^0$ is true, then in the next conjunct all other
actions are blocked, i.e.\ negated, variables $x^1$ and $\ta^1_0$ are updated accordingly. 
Note that these two variables hold the values that are used in the next unfolding.
\qex
\end{example}

\section{Stateful Priorities and Transformation of Networks}
\label{transfo}

This section explains our approach for synthesizing stateful priorities, and how we use them for imposing precise restrictions on system actions.
That explanation requires introducing the following definitions.

In Section~\ref{prelim}, we introduce the notion of state for networks of discrete automata, now we introduce the
analogous notion, \emph{configuration}, for encoded networks. Intuitively, configurations 
can be considered as extended states since they 
%differ from states in that they
hold the same kind of information that states hold, together with 
additional information related to actions and transition steps. We use configurations to store information from 
satisfying assignments output by a $\theory$-satisfiability solver.

\begin{definition}[Configuration] 
Let $\dbs{\dot{\nta}}_k$ be as defined in Definition~\ref{encodingNet}.
A \emph{configuration} $\cnf$ of $\dbs{\dot{\nta}}_k$ is a tuple $\preerror$, 
where $\mathit{loc}: \vars^{\ta} \pto \locs(\nta')$, 
$\mathit{var}: \vars^{\var} \pto \mathbb{Z}\cup\mathbb{R}\cup\mathbb{B}$ and $\mathit{act}: \vars^{\act} \pto \mathbb{B}$
are partial functions respectively mapping variables in $\vars^{\ta}$ to locations of $\nta'$; 
in $\vars^{\var}$ to integer, real and boolean values; 
in $\vars^{\act}$ to boolean values; and $\mathit{stp}$ is an integer variable. 
We write $\mathit{loc}_{\cnf}$, $\mathit{var}_{\cnf}$, $\mathit{act}_{\cnf}$ and $\mathit{stp}_{\cnf}$ to refer to the elements of $\cnf$. 
\qex
\label{configuration}
\end{definition}

Section~\ref{encoding} provides an encoding of networks, however, that is only useful for describing networks to be analysed.
Now, we provide means, for instance, to describe errors to be reached.
Moreover, when using SMT-solving techniques as we do in this work, 
we require to control the reachability analysis at each unfolding step. To this end, 
we introduce new functions that output constraints which are used in our reachability analysis, and synthesis of stateful priorities.

\begin{definition}[Progress, Query, Avoid, PreError and Error Constraints] 
Let $\dbs{\dot{\nta}}_k$ be defined as in Defi\-nition~\ref{encodingNet}.
Let $\cnf$ be a configuration of $\dbs{\dot{\nta}}_k$. 
Let $j\in\nat$.
We encode the following: 
\begin{itemize}
\item \emph{progress} constraints, enforce transitions through unvisited states, i.e.\ \\
	$\textbf{P}(j):=\bigwedge_{i=0}^{j} \bigvee_{\substack{\ta\in\nta,\var\in\vars(\ta)}} (\ta^i \neq \ta^{i+1} \lor \var^i \neq \var^{i+1})$,
\item \emph{query} constraints, encode the error to reach, i.e.\ \\ 
$\textbf{Q}(j,\cnf):=\bigwedge_{\substack{\ta\in\nta,\var\in\vars(\ta)}} \ta^{j+1} = \mathit{loc}_{\cnf}(\ta^{j}) \land \var^{j+1} = \mathit{var}_{\cnf}(\var^{j})$,
\item \emph{avoid} constraints, avoid reaching already reached preErrors, i.e.\ \\
$\textbf{D}(j,\cnf):=\bigwedge_{i=0}^{j}\bigvee_{\substack{\ta\in\nta,\var\in\vars(\ta)}}(\ta^{i} \neq \mathit{loc}_{\cnf}(\ta^{i}) \lor \var^{i} \neq \mathit{var}_{\cnf}(\var^{i}))$,
\item \emph{preError} constraints, encode a preError to synthesize priorities from it, i.e.\ \\
$\textbf{R}(j,\cnf):=\bigwedge_{\substack{\ta\in\nta,\var\in\vars(\ta)}} \ta^{j} = \mathit{loc}_{\cnf}(\ta^{j}) \land \var^{j} = \mathit{var}_{\cnf}(\var^{j})$ and, 
\item \emph{error} constraints, encode the error to avoid by synthesizing priorities, i.e.\ \\
$\textbf{E}(j,\cnf):= \bigvee_{\substack{\ta\in\nta,\var\in\vars(\ta)}} \ta^{j+1}\neq \mathit{loc}_{\cnf}(\ta^{j}) \lor \var^{j+1}\neq\mathit{var}_{\cnf}(\var^{j})$.\qex
\end{itemize}
\label{constraints}
\normalsize
\end{definition}

We introduce the following notion of stateful priority. Intuitively, a stateful priority is a pair consisting of a 
configuration denoting a preError, and
a priority which from that preError restricts an action leading to the error. 

\begin{definition}[Stateful Priority] 
A \emph{stateful priority} is a pair $(\cnf,\rho)$, where $\cnf$ is a configuration wrt.\ a given preError,
and $\rho$ is priority synthesized from $\cnf$. 
$\mathit{Prios}(\nta)$ denotes the set of all stateful priorities wrt.\ $\nta$.
We call a $(\cnf,\rho)\in\mathit{Prios}(\nta)$ \emph{reflexive} if and only if $\rho_{\mathit{be}}=\rho_{\mathit{br}}$. 
We call $(\cnf,\rho)\neq(\bar\cnf,\bar\rho)\in\mathit{Prios}(\nta)$ \emph{circular} if and only if 
$\cnf=\bar\cnf \land (\rho_{\mathit{be}}=\bar\rho_{\mathit{br}} \lor \rho_{\mathit{br}}=\bar\rho_{\mathit{be}})$.
\qex
\label{stateful}
\end{definition}

\iffalse
\begin{definition}[Function $\mathit{actToError}$] 
Let $\dbs{\dot{\nta}}_k$ be as defined in Definition~\ref{encodingNet}.
Let $\act\in\acts(\nta)$ be an action of $\nta$.
Let $\cnf_{\mathit{preErr}}$ be a preError configuration of $\dbs{\dot{\nta}}_k$. 
Let $\mathit{Errors}$ be a set of configurations of $\dbs{\dot{\nta}}_k$ denoting errors.
Then function $\mathit{actToError}$ takes as inputs $\cnf_{\mathit{preErr}}$, $\act$ and $\mathit{Errors}$, and
retrieves true if there exists a $\cnf_{error}\in\mathit{Errors}$ such that $\act$ induces a transition from $\cnf_{\mathit{preErr}}$ to $\cnf_{error}$,
and false otherwise, i.e.\ if
\vspace{-0.5cm}
\[
\mathit{actToError}(\cnf_{\mathit{preErr}},\act,\mathit{Errors})=
\begin{cases}
\mathit{true},  & \text{if \textcolor{white}{-}} 
 		\exists\cnf_{error}\in\mathit{Errors} \ \bullet \cnf_{preError}\leadsto_{\act}\cnf_{error},  \\ 
\mathit{false}, & \text{otherwise,}
\end{cases}
\]
where $\leadsto$ is a transition relation between configurations equivalent to the relation $\rightarrow$ from Section~\ref{prelim}.
\qex
\label{priority}
\end{definition}
\fi

\subsection[]{Synthesis of Stateful Priorities}
% 
%After synthesizing those priorities we assist the user by automatically incorporating them into guards of
%a transformed network, and thereby blocking all transitions that in the original network reached the error. 
%
We now present algorithms for reaching preErrors, and for synthesizing stateful priori\-ties from them.
Given: 
(a) a network $\nta$, 
(b) an error formula $\phi$ and, 
(c) a $\mathit{Max}$ integer, 
those algorithms perform two main tasks:
(1) to compute the set of all reachable preErrors wrt.\ $\phi$ and, 
(2) to synthesize a set of stateful priorities from each preError. 
Algorithm~\ref{algMain} in line $2$ calls function $\mathit{getEncoding}$ on $\nta$ and $\mathit{Max}$, 
for obtaining an encoding as in Definition~\ref{encodingNet}. Function $\mathit{getErrorConfig}$ obtains a configuration
wrt.\ $\phi$. In line $3$, procedure $\mathit{explore}$ is called to perform the above mentioned tasks whose details are
given as follows.

%\subsection[]{Task 1: Reachability of PreErrors}

\textbf{Task 1: Reachability of PreErrors}. In Algorithm~\ref{algExplore}, $\mathit{explore}$ reaches $\cnf_{error}$ stepwise up to $\mathit{Max}$-steps (lines $3$-$7$).
In more detail, $\mathit{checkReach}$ is called to check whether or not $\cnf_{error}$ is reachable at the step $\mathit{cnt}$ (line $4$).
If $\cnf_{error}$ is reachable, then a related preError is also reacha\-ble, and it is collected into the set $\mathit{PreErrors}$, 
otherwise the check is performed with the next step (lines $5$-$6$).
In Algorithm~\ref{algReach}, $\mathit{checkReach}$ calls $\mathit{encodeReachability}$ (line $3$)
for enco\-ding a reacha\-bility problem using $\dbs{\dot{\nta}}_{\mathit{Max}}$ 
conjoined with cons\-traints for the current step output by:
$\textbf{P}(\mathit{step})$, $\textbf{D}(\mathit{step},\cnf_{preError})$ for each $\cnf_{preError}\in\mathit{PreErrors}$ and, 
$\textbf{Q}(\mathit{step},\cnf_{error})$.
The resulting encoding is passed to the function $\theory$-$\mathit{sat}$ (line $4$).
If $\theory$-$\mathit{sat}$ returns a satisfying assignment, then a preError related to $\cnf_{error}$ has been found,
and a respective preError configuration, $\cnf_{preError}$, is created 
(by function $\mathit{createConfig}$ using the satisfying assignment) and collected (line $5$).  

%\subsection[]{Task 2: Synthesis of Priorities}

\textbf{Task 2: Synthesis of Priorities}. In Algorithm~\ref{algExplore}, procedure $\mathit{explore}$ ite\-rates the set $\mathit{PreErrors}$, for synthesizing
a set of stateful priorities from each collected preError (lines $8$-$13$). 
If no priorities for the underlying preError are synthesized by function $\mathit{checkPrios}$ (line $9$), then 
it is considered a new error (given that it unavoidably leads to $\cnf_{error}$),
and the process of reaching preErrors wrt.\ that new error starts over (lines $9$-$12$).
Note that all errors are collected in $\mathit{Errors}$.
Clearly, the process of reaching preErrors is not performed for the initial configuration ($\cnf_{init}$).
In Algorithm~\ref{algPrio}, procedure $\mathit{checkPrios}$ calls $\mathit{encodeSynthesis}$ 
for encoding a synthesis problem using $\dbs{\dot{\nta}}_{\mathit{Max}}$ conjoined with cons\-traints for the current step output by:
$\textbf{P}(\mathit{step})$, 
$\textbf{R}(\mathit{step},\cnf_{preError})$ and, 
$\textbf{E}(\mathit{step},\cnf_{error})$ for each $\cnf_{error}\in\mathit{Errors}$ (line $3$).
The resulting encoding is passed to $\theory$-$\mathit{sat}$ (line $4$).
If $\theory$-$\mathit{sat}$ returns $\perp$, then the preError is considered a new error, otherwise
using the returned satisfying assignment we create a 
configuration $\cnf$, only containing the action that avoids the error expressed by the respective constraint (line $6$).

For synthesizing a priority $\mathit{checkPrios}$ calls $\mathit{createPrio}$ (line $7$). 
This function obtains from $\cnf_{preError}$ an action (the blockee) that reaches an error in $\mathit{Errors}$, 
and from $\cnf$ an action (the blocker) which differs from the blockee, and which from $\cnf_{preError}$ avoids that error.
Function $\mathit{checkCircular}$ checks that a newly synthesized stateful priority is not a circular one (line $8$). 
To this end, this function iterates the set $\mathit{Stateful}$ in order to find circular stateful priorities wrt. the newly synthesized one.
If circular stateful priorities are synthesized from the underlying preError, then it is considered a new error, otherwise
the stateful priority is collected (line $9$). 
Note that we use the same blockee for synthesizing fresh stateful priorities with new blockers (lines $10$-$11$), if any.
 
\begin{algorithm}[t]
\caption{Main Procedure}\label{algMain}
\begin{algorithmic}[1]
\scriptsize 
\Procedure{}{}$\mathit{main}(\nta,\phi,\mathit{Max})$ 
%\hspace{-15cm}
\State $\dbs{\dot{\nta}}_{\mathit{Max}}:= \mathit{getEncoding}(\nta,\mathit{Max})$; $\cnf_{error}:=\mathit{getErrorConfig}(\phi)$;
	   $\mathit{Errors}:=\{\cnf_{error}\}$; $\mathit{Stateful}:=\emptyset$; \text{   // Global variables }
\State $\mathit{explore}(\cnf_{error})$;
\EndProcedure
\end{algorithmic}
\end{algorithm}

\begin{algorithm}[t]
\caption{Exploring States of Network}\label{algExplore}
\begin{algorithmic}[1]
\scriptsize 
\Procedure{}{}$\mathit{explore}(\cnf_{error})$
%\hspace{-15cm}
\State $\mathit{cnt}:=0$; $\mathit{PreErrors}:=\emptyset$; $\mathit{PEs}:=\emptyset$;
\While {$\mathit{cnt} < \mathit{Max}$}  
\State $\mathit{PEs}:= \mathit{checkReach}(\mathit{PreErrors},\cnf_{error},\mathit{cnt})$;
\If {$\mathit{PEs}\neq\emptyset$} $\mathit{PreErrors}:=\mathit{PreErrors}\cup\mathit{PEs}$; $\mathit{cnt}:=0$; \Else \ $\mathit{cnt}:=\mathit{cnt}+ 1$;
\textbf{end} \EndIf 
\EndWhile
\State \textbf{end}
\For {\textbf{each}  $\cnf\in\mathit{PreErrors}$} 
\If {$\mathit{checkPrios}(\cnf)=\mathit{false}$ \textbf{and} $\cnf\neq\cnf_{init}$}  
\State $\mathit{Errors}:= \mathit{Errors} \cup \{\cnf\}$;
\State $\mathit{explore}(\cnf)$;
\EndIf
\State \textbf{end}
\EndFor
\State \textbf{end}
\EndProcedure
\end{algorithmic}
\end{algorithm}

\begin{algorithm}[h!]
\caption{Checking Reachability}\label{algReach}
\begin{algorithmic}[1]
\scriptsize 
\Procedure{}{}$\mathit{checkReach}(\mathit{PreErrors},\cnf_{error},\mathit{step})$
\State $\mathit{PEs}:=\emptyset$; 
\State $\mathit{Net}:= \mathit{encodeReachability}(\mathit{PreErrors},\cnf_{error},\mathit{step})$; 
\State $\assig_{\mathit{sat}}:= \theory$-$\mathit{sat}(\mathit{Net})$;
\If {$\assig_{\mathit{sat}} \neq \perp$} $\cnf_{preError}:=\mathit{createConfig}(\assig_{\mathit{sat}})$; $\mathit{PEs}:=\{\cnf_{preError}\}$;  \textbf{end}
\EndIf
\EndProcedure
\State \textbf{return} $\mathit{PEs}$;
\end{algorithmic}
\end{algorithm} 

\iffalse
\begin{algorithm}[t]
\caption{Checking Reachability}\label{algReach}
\begin{algorithmic}[1]
\scriptsize 
\Procedure{}{}$\mathit{checkReach}(\mathit{PreErrors},\cnf_{error},\mathit{step})$
\State $\mathit{PEs}:=\emptyset$; 
 $\mathit{Net}:= \dbs{\dot{\nta}}_{\mathit{Max}} \land \textbf{P}(\mathit{step})$;
\For {\textbf{each} $\cnf_{pe}\in\mathit{PreErrors}\cup\mathit{Errors}$} 
$\mathit{Net}:= \mathit{Net} \land \textbf{D}(\mathit{step},\cnf_{pe})$; \textbf{end}  
\EndFor
\State $\mathit{Net}:= \mathit{Net}\land \textbf{Q}(\mathit{step},\cnf_{error})$; 
\State $\assig_{\mathit{sat}}:= \theory$-$\mathit{sat}(\mathit{Net})$;
\If {$\assig_{\mathit{sat}} \neq \perp$} 
\State $\cnf:= \Big(\substack{\{\var\mapsto\dbs{\var}_{\assig_{\mathit{sat}}}\mid\var\in\varslocs{\mathit{Net}}{\mathit{step}}\},
\{\var\mapsto\dbs{\var}_{\assig_{\mathit{sat}}}\mid\var\in\varsvars{\mathit{Net}}{\mathit{step}}\},
\{\var\mapsto\dbs{\var}_{\assig_{\mathit{sat}}}\mid\var\in\varsacts{\mathit{Net}}{\mathit{step}}\}, \mathit{step}}\Big)$; $\mathit{PEs}:=\{\cnf\}$; 
\EndIf
\State \textbf{end}
\EndProcedure
\State \textbf{return} $\mathit{PEs}$;
\end{algorithmic}
\end{algorithm} 
\fi

\begin{algorithm}[h!]
\caption{Checking Priorities}\label{algPrio}
\begin{algorithmic}[1]
\scriptsize 
\Procedure{}{}$\mathit{checkPrios}(\cnf_{preError})$
\State  $\mathit{found}:=\mathit{false}$; $\mathit{step}:= \mathit{stp}_{\cnf_{preError}}$;
\State $\mathit{Net}:= \mathit{encodeSynthesis}(\cnf_{preError},\mathit{step})$; 
\State $\assig_{\mathit{sat}}:= \theory$-$\mathit{sat}(\mathit{Net})$;
\While {$\assig_{\mathit{sat}} \neq \perp$}   
\State $\cnf:=\mathit{createConfig}(\assig_{\mathit{sat}})$;
\State $\rho:=\mathit{createPrio}(\cnf_{\mathit{preError}},\cnf)$;
\If {$\mathit{checkCircular}(\cnf_{preError},\rho)$}  $\mathit{found}:=\mathit{false}$; \textbf{break};
\Else \ $\mathit{Stateful}:= \mathit{Stateful}\cup\{(\cnf_{preError},\rho)\}$; $\mathit{found}:=\mathit{true}$;  \textbf{end}
\EndIf
\State $\mathit{Net}:= \mathit{Net} \land \neg\rho_{br}^{\mathit{step}}$;
\State  $\assig_{\mathit{sat}}:= \theory$-$\mathit{sat}(\mathit{Net})$;
\EndWhile
\State \textbf{end}
\EndProcedure
\State \textbf{return} $\mathit{found}$;
\end{algorithmic}
\end{algorithm}

\subsection[]{Transformation of Networks}
\label{trafoNet}

We present function~$\repifygrd$ and algorithm~$\alg^{\rho}$ for using synthesized stateful priorities.
$\alg^{\rho}$ introduces in guards and update vectors of edges of networks positional variables, which
hold the current locations of all automata of a network at a particular state. 
For blocking transitions to a given error state, 
we use positional variables in guards of edges whose actions appear as blockees in stateful prio\-rities.
In other words, positional variables encode preError states from which transitions 
that exclusively avoid reaching that error state are induced.   

\begin{definition}[Function $\repifygrd$] 
Let $\nta = \{\ta_1, \dots, \ta_N\}$ be a network.
Let $p_{\ta_1},\dots,$\\$p_{\ta_N}$ be integer positional variables.
Let $k\in\natplus$.
Let $\dbs{\dot{\nta}}_k$ be $\nta$ encoded as in Definition~\ref{encodingNet}, from which 
$\dot\vars$ is the set of varia\-bles interpreted over $\locs(\nta)$ used in $\dbs{\dot{\nta}}_k$.
Let $\mathit{SP}$ be the set of stateful priorities obtained in Algorithms~\ref{algMain}-\ref{algPrio}.
Using $\loc(\edge)$, $\loc'(\edge)$ and $\act(\edge)$ to
denote the source, destination, and action of an edge $\edge$, then $\repifygrd(\constr,\edge,\mathit{SP})=$
%\vspace{-0.31cm}
\small
\[
\begin{cases}
\constr \land \hspace{-0.5cm}\bigvee\limits_{\substack{\ta\in\nta,\\ 0\le i\le k, \ \ta^i\in\dot\vars}} 
		\hspace{-0.5cm}p_{\ta}\neq\mathit{loc}_{\cnf}(\ta^i),  & \text{if \textcolor{white}{-}} 
 		\substack{\exists \rho,\dot\var \ \bullet \ (\cnf,\rho)\in\mathit{SP} 
				\ \land \ \dot\var\in\dot\vars \ \land \ \act(\edge)=\rho_{\mathit{be}} \ \land \\
				\loc(\edge)=\mathit{loc}_{\cnf}(\dot\var) \ \land 
				\loc'(\edge)\neq\mathit{loc}_{\cnf}(\dot\var) \ \land \ 
				\neg\exists\dot\edge\in\edges(\nta) \ \bullet \\ \loc(\edge)=\loc(\dot\edge) 
				\ \land \ \loc'(\edge)\neq\loc'(\dot\edge) \ \land \
				\act(\dot\edge)=\rho_{\mathit{br}},}  \\ 
\constr \land \big(\hspace{-0.65cm}\bigvee\limits_{\substack{\ta\in\nta,\\ 0\le i\le k,\ \ta^i\in\dot\vars}} 
		\hspace{-0.5cm}p_{\ta}\neq\mathit{loc}_{\cnf}(\ta^i)\big)\land p_{\ta(\edge)} \neq \loc(\edge),  & \text{if \textcolor{white}{-}} 
 		\substack{\exists \rho,\dot\var \ \bullet \ (\cnf,\rho)\in\mathit{SP} 
				\ \land \ \dot\var\in\dot\vars \ \land \ \act(\edge)=\rho_{\mathit{be}} \ \land \\
				\loc(\edge)=\mathit{loc}_{\cnf}(\dot\var) \ \land \ 
				\loc'(\edge)\neq\mathit{loc}_{\cnf}(\dot\var) \ \land \ 
				\exists\dot\edge\in\edges(\nta) \ \bullet \\ \loc(\edge)=\loc(\dot\edge) 
				\ \land \ \loc'(\edge)\neq\loc'(\dot\edge) \ \land \
				\act(\dot\edge)=\rho_{\mathit{br}},}  \\ 
\constr, & \text{otherwise.}
\end{cases} 
\]
\normalsize  
\label{omg}
\end{definition}

Note that the second condition outputs a more restrictive guard than the first condition. 
That guard includes an extra conjunct for blocking computation paths that unavoidably lead to an error state. For instance, 
assume that for $\nta_1$ any state where $\ta_0.3$ is reached denotes an error. 
We would use this more restrictive guard for blocking any transition with $a$ whenever $\ta_0$ is located at $1$, 
since from this location a transition with $a$ unavoidably leads to $\ta_0.3$. 

\begin{definition}[Transformation Algorithm~$\alg^{\rho}$] 
Let $\nta$ be a network. 
Let $k\in\natplus$.
Let $\dbs{\dot{\nta}}_k$ be $\nta$ encoded as in Definition~\ref{encodingNet}.
Let $\mathit{SP}$ be the set of stateful priorities obtained in Algorithms~\ref{algMain}-\ref{algPrio}.
The output of $\algprio$ is $\nta^{\rho}=\{\algprio(\ta,\mathit{SP})\mid \ta\in\nta\}$, 
with $\algprio(\ta,\mathit{SP})= (\locs(\ta),\acts(\ta),\vars',\edges',\iloc{})$ where:
$\vars'= \vars(\ta)\cup\{\posvar{\ta}\mid \ta\in\nta\land \mathit{SP}\neq\emptyset\}$, i.e.\ a fresh integer positional varia\-ble for each 
$\ta\in\nta$ is added (initial value is $\iloc{,\ta}$), and 
$\edges'=\{(\loc,\act,\repifygrd(\constr,\edge,\mathit{SP}),\langle\var_1:=\expr[1],\dots,\var_m:=\expr[m],\posvar{\ta}:=\loc'\rangle,\loc')
\mid\edge=(\loc,\act,\constr,\langle\var_1:=\expr[1],\dots,\var_m:=\expr[m]\rangle,\loc')\in\edges(\ta)\}$.
\qex
\label{trafo}
\end{definition}

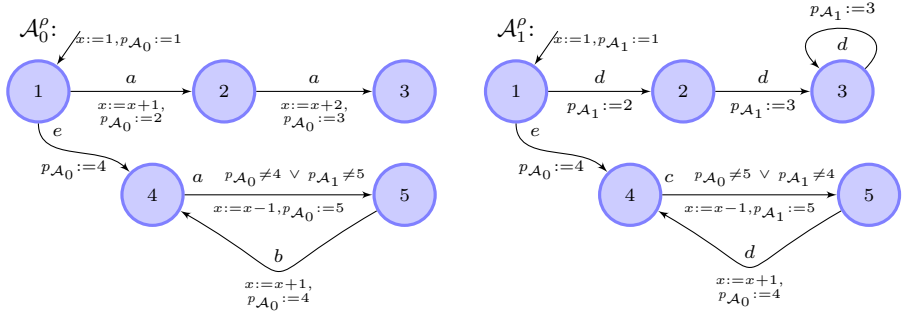
\begin{figure}[t!]
\centering
\begin{tikzpicture}[>=latex',join=bevel,]
\tikzstyle{every state}=     [draw=blue!50,very thick,fill=blue!20]%
\tikzstyle{state2}=[draw=white,fill=white]%

\node (A0) at (55bp,113bp) [state2] {};
\node (A1) at (37bp,87bp) [state] {\scriptsize $1$};
\node (A2) at (107bp,87bp) [state] {\scriptsize $2$};
\node (A3) at (175bp,87bp) [state] {\scriptsize $3$};
\node (A4) at (80bp,48bp) [state] {\scriptsize $4$};
\node (A5) at (175bp,48bp) [state] {\scriptsize $5$};
\node (F2) at (37bp, 110bp)  {$\ta_0^{\rho}$:};

\draw [->] (A0) to[]  node[above] {}  node[below,xshift=0.80cm, yshift=0.30cm] {\scriptsize $\substack{\clk:=1, \posvar{\ta_0}:=1}$}(A1); 
\draw [->] (A1) to[]  node[above] {\scriptsize $a$}  node[below] {\scriptsize $\substack{\clk:=x+1,\\\posvar{\ta_0}:=2}$}(A2); 
\draw [->] (A2) to[]  node[above] {\scriptsize $a$}  node[below] {\scriptsize $\substack{\clk:=x+2,\\\posvar{\ta_0}:=3}$}(A3);
\draw [->] (A1) to[out=270]  node[above, xshift=-0.21cm, yshift=0.1cm] {\scriptsize $e$}  node[below] 
{$\substack{\posvar{\ta_0}:=4}$}(A4);
\draw [->] (A4) to[]  node[above] {\scriptsize $a\text{\textcolor{white}{---}} \substack{\posvar{\ta_0}\neq 4 \ \lor \ \posvar{\ta_1}\neq 5}$}  
node[below] {\scriptsize $\substack{\clk:=x-1, \posvar{\ta_0}:=5}$}(A5);
\draw [->] (A5) to[bend left, looseness=2.1]  node[above] {\scriptsize $b$}  node[below] 
{\scriptsize $\substack{\clk:=x+1,\\ \posvar{\ta_0}:=4}$}(A4);

\node (B0) at (235bp,113bp) [state2] {};
\node (B1) at (217bp,87bp) [state] {\scriptsize $1$};
\node (B2) at (280bp,87bp) [state] {\scriptsize $2$};
\node (B3) at (340bp,87bp) [state] {\scriptsize $3$};
\node (B4) at (260bp,48bp) [state] {\scriptsize $4$};
\node (B5) at (350bp,48bp) [state] {\scriptsize $5$};
\node (G2) at (217bp, 110bp)  {$\ta_1^{\rho}$:};

\draw [->] (B0) to[]  node[above] {}  node[below,xshift=0.80cm, yshift=0.30cm] {\scriptsize $\substack{\clk:=1, \posvar{\ta_1}:=1}$}(B1); 
\draw [->] (B1) to[]  node[above] {\scriptsize $d$}  node[below] {$\substack{\posvar{\ta_1}:=2}$}(B2); 
\draw [->] (B2) to[]  node[above] {\scriptsize $d$}  node[below] {$\substack{\posvar{\ta_1}:=3}$}(B3);
\draw [->] (B1) to[out=270]  node[above, xshift=-0.21cm, yshift=0.1cm] {\scriptsize $e$}  
node[below] {$\substack{\posvar{\ta_0}:=4}$}(B4);
\draw [->] (B4) to[]  node[above] {\scriptsize $c\text{\textcolor{white}{---}} \substack{\posvar{\ta_0}\neq5 \ \lor \ \posvar{\ta_1}\neq 4}$}  
node[below] {\scriptsize $\substack{\clk:=x-1, \posvar{\ta_1}:=5}$}(B5);
\draw [->] (B5) to[bend left, looseness=2.1]  node[above] {\scriptsize $d$}  
node[below] {\scriptsize $\substack{\clk:=x+1,\\ \posvar{\ta_0}:=4}$}(B4);
\draw [->] (B3) to[loop, looseness=1, loop, distance=1cm]  node[above] {$\substack{\posvar{\ta_1}:=3}$} 
 node[below] {\scriptsize $d$}(B3);
\end{tikzpicture}
\vspace*{-1em}
\caption{Network $\nta^{\rho}_1$ using positional variables to incorporate synthesized priorities.}
\label{fig2}
\end{figure}

\begin{example}
Figure~\ref{fig2} shows network $\nta^{\rho}_1$ 
obtained from $\nta_1$ by introducing the positional varia\-bles $\posvar{\ta_0}$ and 
$\posvar{\ta_1}$ in guards and update vectors of edges.
Applying Algorithms~\ref{algMain}-\ref{algPrio} on $\nta_1$ 
we obtain (after some simplifications)
$\mathit{Stateful}=$\small $\{(\cnf:=(\mathit{loc}(\ta_0)=4,\mathit{loc}(\ta_1)=5,\mathit{var}(x)=0),\rho:=(a,d)), 
(\dot\cnf:=(\mathit{loc}(\ta_0)=5,\mathit{loc}(\ta_1)=4,\mathit{var}(x)=0),\dot\rho:=(c,b))\}$\normalsize. 
We use each pair in $\mathit{Stateful}$ for extending guards of edges whose actions appear as blockees in priorities.
For instance, by using $(\cnf,\rho)$ function $\Gamma$ extends the guard of 
$\edge_1=(4,a,\true,\langle x:=x-1\rangle,5)$, because: 
(1) its action is a blockee in $\rho$, i.e.\ $\act(\edge_1)=\rho_{be}$, (2) its origin is where $\ta_0$ is located in $\cnf$, 
i.e.\ $\loc(\edge_1) = \mathit{loc}_{\cnf}(\ta_0)$ and, (3) its destination is not where $\ta_0$ is located in $\cnf$,
i.e.\ $\loc'(\edge_1) \neq \mathit{loc}_{\cnf}(\ta_0)$. In this extension, $\posvar{\ta_0}$ and $\posvar{\ta_1}$ are used 
in a disjunction to differ from $\mathit{loc}_{\cnf}(\ta_0)$ and  $\mathit{loc}_{\cnf}(\ta_1)$, respectively. 
Similarly, the guard of $\edge_2=(4,c,\true,\langle x:=x-1\rangle,5)$ is extended by using $(\dot\cnf,\dot\rho)$.
Note that the edges modified by $\Gamma$ exclusively induce transitions that avoid reaching $\phi$. For ins\-tance, 
a transition with $c$ is blocked whenever $\ta_0$ and $\ta_1$ are respectively located at $5$ and $4$. Similarly,
a transition with $a$ is blocked whenever $\ta_0$ and $\ta_1$ are respectively located at $4$ and $5$.
Note that other transitions remain unrestricted in the transformed network $\nta^{\rho}_1$.
\qex
\end{example}

The following lemma shows that our approach does not introduce deadlocks in transformed networks.

\begin{lemma}
Let $\nta$ be a network, and $\phi$ an error formula such that $\nta\models\epfq{\phi}$. 
Let $\mathit{SP}$ be the set of stateful priorities 
obtained by using Algorithms~\ref{algMain}-\ref{algPrio} on $\nta$,
$\phi$ and on a $k\in\natplus$ which is big enough to reach $\phi$. Let $\nta^{\rho}=\{\algprio(\ta,\mathit{SP})\mid \ta\in\nta\}$. 
Then, $\algprio$ does not introduce (new) deadlocks.
\qex
\label{theorem1}
\end{lemma}
\textbf{Proof}. Assume a set 
$P= \{(\rho_1,\dots,\rho_m)\}$ of synthesized priorities wrt.\ $\nta$ and $\phi$.
Priorities in $P$ induce deadlocks in $\nta^{\rho}$ in the following cases:
\begin{enumerate}
\item 
Transitions are blocked by reflexive priorities or by circular priorities.
In Algorithm~\ref{algPrio}, line $7$ function $\mathit{createPrio}$ synthesizes non-reflexive priorities; 
procedure $\mathit{checkCircular}$ in line $8$ avoids synthesizing circular priorities.
\item Transitions not leading to $\phi$ are blocked by some priorities in $P$.
Let $\cnf_{error}$ be the error configu\-ration denoted by $\phi$.
Let $\cnf^1_{preError},\dots,\cnf^t_{preError}$, $t\ge 0$, be preErrors configurations related to $\cnf_{error}$, 
which by not yielding priorities became the new errors to be reached (Algorithm~\ref{algExplore}, lines $10$-$11$).
Pick a configuration $\cnf$ from which edges $\edge_1,\dots\edge_j$, 
induce a transition on action $\act$ that does not lead, neither to $\cnf_{error}$, nor to any $\cnf^i_{preError}$, $1\le i\le t$.
Then, function $\mathit{createPrio}$ does not uses action $\act$ as blockee in a stateless priority (Algorithm~\ref{algPrio}, line $7$).
Thus, $\algprio$ (through $\Gamma$) does not extend the guards of edges $\edge_1,\dots,\edge_j$.
Thus, edges $\edge_1,\dots,\edge_j$ remain unchanged in $\nta^{\rho}$.
\end{enumerate}
\vspace{-0.18cm}
Hence, priorities in $P$ are not synthesized by our approach.
\textcolor{white}{-------------------------}
$\qed$\\

Note that using positional variables in transformed networks does not introduce any additional beha\-vior, 
at the contrary, those variables restrict behavior. This is clear from the fact that 
positional variables just store the current locations of automata in a given state, thus no new information is introduced.

In the following, Lemma~\ref{lemmaformula} presents a formula to calculate an upper bound on the number of reachable states in transformed networks.
This lemma uses the observation that Algorithm~\ref{algExplore} collects in the set $\mathit{Errors}$, 
states which in transformed networks become unreachable. However, not every unreachable state 
is collected in that set. Note that transformed networks may use the most restrictive transformation of guards from Definition~\ref{omg}, which 
restricts states not collected in $\mathit{Errors}$ from being reachable.

\begin{lemma}
Let $\nta$ be a network, and $\phi$ an error formula such that $\nta\models\epfq{\phi}$. 
Let $\mathit{SP}$ be the set of stateful priorities 
obtained by using Algorithms~\ref{algMain}-\ref{algPrio} on $\nta$,
$\phi$ and on a $k\in\natplus$ which is big enough to reach $\phi$. 
Let $\mathit{Errors}$ be the set of collected errors in Algorithm~\ref{algExplore}.
Let $\nta^{\rho}=\{\algprio(\ta,\mathit{SP})\mid \ta\in\nta\}$. 
Then the number of reachable configurations in $\nta^{\rho}$ is bounded above by
$|\reachconf{\nta}| - |\mathit{Errors}|$.
\qex
\label{lemmaformula}
\end{lemma}

\begin{corollary} The number of reachable states in transformed networks is in the worst case, the same as in their original counterparts, i.e.\ 
$|\reachconf{\nta^{\rho}}|\leq|\reachconf{\nta}|$. 
\qex
\end{corollary}

\section{Correctness of Transformation}
\label{correctness}

In this section, we discuss the correctness of our transformation approach. 
We set that discussion by considering the following example.
 
\begin{figure}[t]
\centering
\begin{tikzpicture}[>=latex',join=bevel,]
\tikzstyle{every state}=[draw=blue!50,very thick,fill=blue!20]%
\tikzstyle{state2}=[draw=white,fill=white]%

\node (A0) at (60bp,110bp) [state2] {};
\node (A1) at (37bp,87bp) [state] {\scriptsize $1$};
\node (A2) at (100bp,87bp) [state] {\scriptsize $2$};
\node (F2) at (37bp, 109bp)  {$\ta_0$:};

\draw [->] (A0) to[]  node[above] {}  node[below,xshift=0.65cm, yshift=0.25cm] {}(A1); 
\draw [->] (A1) to[]  node[above] {\scriptsize $a$}  node[below] {}(A2); 
\draw [->] (A2) to[bend left]  node[above] {\scriptsize $a$}  node[below] {}(A1); 

\node (B0) at (240bp,110bp) [state2] {};
\node (B1) at (217bp,87bp) [state] {\scriptsize $1$};
\node (B2) at (280bp,87bp) [state] {\scriptsize $2$};
\node (G2) at (217bp, 109bp)  {$\ta_1$:};

\draw [->] (B0) to[]  node[above] {}  node[below,xshift=0.65cm, yshift=0.25cm] {}(B1); 
\draw [->] (B1) to[]  node[above] {\scriptsize $b$}  node[below] {}(B2);
\draw [->] (B1) to[bend left]  node[above] {\scriptsize $c$}  node[below] {}(B2);
\draw [->] (B2) to[bend left]  node[above] {\scriptsize $b$}  node[below] {}(B1);
\end{tikzpicture}

\vspace*{-1em}
\caption{Network $\nta_2$ of discrete automata.}
\label{fig3}
\end{figure}
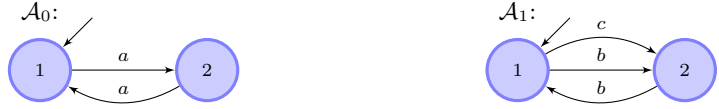

\begin{example}
For the network $\nta_2$ in Figure~\ref{fig3}, consider the error $\phi:=\ta_0.2 \land \ta_1.2$. 
The following preErrors are reachable: $\tsconf = \langle(\ta_0.1,\ta_1.2)\rangle$ and 
$\dot\tsconf=\langle(\ta_0.2,\ta_1.1)\rangle$. The following priorities are synthesized: $\{(a,b)\}$ from $\tsconf$, 
and $\{(b,a), (c,a)\}$ from $\dot\tsconf$. Note that our definition of circularity for stateful priorities is less restrictive than 
the one often used for stateless priorities (see for instance~\cite{Bornot,Cheng,Cheng2}). 
Considering stateless priorities each priority in the set $\{(a,b)\}\cup\{(b,a), (c,a)\}$ is a circular one, thus, there is no way to avoid $\phi$, 
since each action in $\nta_2$ is blocked by another action. 
However, in this work circularity is defined on stateful priorities. 
That is, priorities are circular only wrt.\ the same related preError.
For instance, $(a,b)$ and $(b,a)$ are not circular because they are related to different preErrors. Thus, 
action $b$ is not preferred over action $a$, and $a$ is not preferred over $b$ wrt.\ the same preError.
Thus, by using stateful priorities (as constructed by algorithm $\nta^{\rho}$) 
we are indeed able to avoid reaching the state denoted by $\phi$.
\qex
\end{example}
 
Admittedly, it is still possible to reach a given error after applying our network transformation, 
if a pair of synthesized stateful priorities are circular. In this case, those priorities are ruled out,
and the error is reached by taking the actions which in those priorities appear as blockees.

Circular priorities are synthesized from 
transitions outgoing from preErrors which are justified by the the same 
action that reaches and avoids a given error. In this way, 
the same action becomes blockee and blocker in different priorities.
The following semantical restriction avoids circular priorities. 
This restriction guarantees that when stateful priorities are synthesized, 
then the underlying error becomes unreachable in a transformed network that uses those priorities.

\begin{definition}[Semantical Restriction for Avoiding Circular Priorities]
Let $\nta$ be a network. 
Let $\tsconf$ be an error state reachable in $\nta$. 
A semantical restriction avoids circular priorities, 
if there does not exist an action that at the same time reaches and avoids $\tsconf$, 
from a preError wrt.\ $\tsconf$, i.e.\  
$\neg\exists\comp,\bar\comp\in\paths{\nta}\bullet
\comp=
\tsconf_0\trans{\act_1}\cdots\trans{\act_{n-1}}\tsconf_{n-1}
\trans{\act_n}\tsconf_n=\tsconf \land
\bar\comp=
\bar\tsconf_0\trans{\bar\act_1}\cdots\trans{\bar\act_{m-1}}\bar\tsconf_{m-1}
\trans{\bar\act_m}\bar\tsconf_m=\tsconf\land
\exists\bar\tsconf_{m-1} \trans{\act_n}\bar\tsconf\bullet \bar\tsconf\neq\tsconf$. 
\qex
\label{semantical}
\end{definition}

Although the above semantical restriction avoids circular priorities,  
it requires model checking the underlying network, in order to know whether or not
circular priorities will be synthesized. With the same objective, 
the following syntactical restrictions can be used a priori.

\begin{definition}[Syntactical Restrictions for Avoiding Circular Priorities]
Let $\nta=\{\ta_1,\dots\ta_N\}$ be a network. 
The following syntactical restrictions avoid circular priorities if and only if in $\nta$:
(1) all automata have disjoint sets of edges, i.e.\
$\acts(\ta_1)\cap\cdots\cap\acts(\ta_N)=\emptyset$ and,
(2) all edges use different actions, i.e.\
$\neg\exists\edge_1=\edgetuple[1]\neq\edge_2=\edgetuple[2]\in\edges(\nta)\bullet\act_1=\act_2$.
\qex
\end{definition}

Note that the above restriction, as opposed to the semantical one, is cheaper to check, 
however, it restricts broadcast transitions completely. 
For our experi\-ments, which perform broadcast transitions, we use the semantical 
restriction to guarantee unreachability of errors because those benchmarks are well known to us, 
and we are sure that the semantical restriction holds in each of those benchmarks without model check them.
One could use the syntactical restrictions, and still have broadcast transitions in a way, 
where broadcast transitions are replaced by atomic ones. That is,
a broadcast transition is modeled as a sequence of uninterrupted action transitions from all
automata participating in the broadcast.

We show in the remaining part of this section
that errors reachable in original systems are unreachable in transformed ones. 

\begin{theorem}[Unreachability of Errors]
Let $\nta$ be a network, and $\phi$ be an error formula such that $\nta\models\epfq{\phi}$, 
and such that $\nta$ fulfills Definition~\ref{semantical} for each reachable preError wrt.\ $\phi$. 
Let $\mathit{SP}$ be the set of stateful priorities 
obtained by using Algorithms~\ref{algMain}-\ref{algPrio} on $\nta$,
$\phi$ and on a $k\in\natplus$ which is big enough to reach $\phi$. 
Let $\nta^{\rho}=\{\algprio(\ta,\mathit{SP})\mid \ta\in\nta\}$. 
Then, $\mathit{SP}\neq\emptyset\Leftrightarrow\nta^{\rho}\not\models\phi$.
\end{theorem}
\textbf{Proof}. (Only $\Rightarrow$, $\Leftarrow$ is trivial). 
Let $\tsconf$ be a state denoting $\phi$.
Given that $\nta$ fulfills Definition~\ref{semantical} for each reacha\-ble preError wrt.\ $\tsconf$, 
thus, no circular priorities are synthesized from each preError. 
Thus, if $\mathit{SP}\neq\emptyset$ then $\mathit{SP}$ contains all stateful priorities that avoid reaching $\tsconf$.
Then, $\algprio$ constructs from $\nta$ and $\mathit{SP}$ a network $\nta^{\rho}$,
which using positional variables encodes in guards each preError occurring in $\mathit{SP}$.
Hence, $\nta^{\rho}$ restricts all transitions to $\tsconf$. Thus, $\tsconf$ is unreachable in  $\nta^{\rho}$.
%\textcolor{white}{---------------------------}
\qed

\begin{table}[b]
\centering
\setlength{\tabcolsep}{4pt}
%\raisebox{3ex}{%
\begin{tabular}[b]{ | l r r r r | l r r r r | l r r r r |}
\hline\hline
\multicolumn{1}{l}{ Net} &  \multicolumn{1}{c}{P} & \multicolumn{1}{c}{M} &\multicolumn{1}{c}{ $t(s)$} & \multicolumn{1}{c|}{U} &
\multicolumn{1}{c}{ Net} &  \multicolumn{1}{c}{P} & \multicolumn{1}{c}{M} &\multicolumn{1}{c}{ $t(s)$} & \multicolumn{1}{c|}{U} &
\multicolumn{1}{c}{ Net} &  \multicolumn{1}{c}{P} & \multicolumn{1}{c}{M} &\multicolumn{1}{c}{ $t(s)$} & \multicolumn{1}{c|}{U} 
\\[0.5ex]
\hline
 \textit{R}-$2$  & 4& 0.51 & 70.5 & 15
 &\textit{P}-$2$  & 44 & 0.92 & 1,380.6 & 15
 &\textit{G}-$3$  &  2& 0.45 & 10.3 & 9\\
 \textit{R}-$3$ &  15& 0.59& 1,240.0 & 25
 &\textit{P}-$3$ &  --&  --& -- & 30
 &\textit{G}-$4$ &  9&  1.59& 139.4 & 12\\
 \textit{R}-$4$  &  --& --& -- & 35
 &\textit{}  &  &  &  &
 &\textit{G}-$5$  & --& --& -- & 15	 \\
\hline
\end{tabular}
\caption{
Row `Net' gives X-$N$, that is, benchmark X with $N$ components.
`P' gives the number of stateful priorities synthesized.
`M' gives memory usage in GB,
`$t(s)$' synthesis time in seconds and,
`U' gives the number of maximal unfolding steps used.
Each of our network was transformed in less than 2 seconds.
Experimental environment: Intel~i3, 2.3GHz, 3GB. Ubuntu~11.04.
$Z3$ version 4.4.0.
}
\label{Table1}
\end{table} 

\section{Experimental Results}
\label{exp}

In this section, we show the applicability of our approach. To this end, 
we programmed our algorithms in \emph{Java}, and obtained a prototype tool called CrEStO.
We use $Z3$ for cons\-traint solving. 
Our tool CrEStO synthesizes stateful priorities from 
three real-world networks of discrete automata, namely, 
\textit{R}, \textit{P}~\cite{Skou}, and \textit{G}~\cite{Vechev}.
Note that the last two networks are untimed versions of the original timed ones.
None of our examples yielded circular priorities. This allowed us to 
obtain in each example all stateful priorities, that exclusively restrict 
transitions leading to the respective verified error.
Consider a more detailed description of each example, and of the respective verified error as follows.

In the context of the large German project, 
\emph{Collaborative Embedded Systems (CrESt)}, which involves a consortium of 
more than $20$ companies, universities and research institutions,
\textit{R} addresses a problem from one participating company which is
related to the deployment of transport robots in factories. Often,
in those factories exist narrow areas where at most one robot at a time is allo\-wed
to transit them. Although this restriction avoids crashes among robots, it often leads to
bottle necks, and thereby to delays in the  transportation of goods. 
\textit{R} models a network with $N$ robots and $4$ more components. These models include
boolean, real and integer variables. The biggest component, the robot one, has $5$ locations
and $12$ edges.
We verified the formula \emph{crash}, which states that a crash occurs when more than
one robot (regardless of the direction) transit though those areas.  
A crash is represented in the network by a location of each robot.
CrEStO synthesizes stateful priorities that orchestrate the transit of those areas without
leading to a crash. 

\textit{P} models a CSMA/CD protocol with $N$ slaves, one master and $3$ more components. 
These models include
boolean and integer variables. The biggest component, has $5$ locations
and $6$ edges.
%This is our biggest example which consists of six automata in total.
We verified the formula \emph{collision}, which states that a co\-llision 
occurs when more than one slave at the same time send data to the master. 
A collision is represented in the network by a location of the master.
CrEStO synthesizes stateful priorities that orchestrate the sending of data to the master without
leading to a collision. 

\textit{G} models a non-trivial program of $N$ process execu\-ting in parallel different statements.
The results of execu\-ting those statements are stored in different variables.
Different interleavings of statements executed by these processes may lead either 
to different, or to the same values for two target variables. 
These models include
boolean and integer variables.
The biggest component, has $5$ locations
and $7$ edges.
We verified the formula \emph{value}, which states that
the value of those target variables is the same. 
Both cases, target variables differ on their values, and target variables have the same value,
are represented by dedicated locations of the network. 
CrEStO synthesizes stateful priorities that orchestrate the execution of those statements 
leading to different values for target variables.
 
Table~\ref{Table1} gives figures for our experiments. Rows without results indicate the 
smallest instances of a case study that ran out of memory.
From that table, admittedly, we can observe that our experiments do not scale very well,
however, this does not invalidate the applicability of our approach on real-world networks, 
which is our goal for these experiments.
The reason for those scalability issues can be our implementation, 
given that each time we encode a network either for reachability of preErrors (Algorithm~\ref{algReach}), 
or for synthesis of stateful priorities (Algorithm~\ref{algPrio}), we write 
that encoding into a file, and then we call $Z3$ for solving. 
This definitely creates an unnecessary overhead in our experiments.
Although we use the $Z3$ Java-API for constructing those encodings, and we could solve
directly from that API, we noticed discrepancies on expected satisfying assignments.
Therefore, we preferred the alternative of writing to file, since the results of
this alternative match our expectations. 
Definitely, we will avoid writing to file as explained before  
by using other APIs.

\section{Related Work} 
\label{related}

\emph{Priority systems}~\cite{Gossler,Gossler2,Gossler3} use priorities 
to represent res\-trictions of behavior of systems. Those restrictions are induced by deadlock-free controllers 
which preserve safety proper\-ties of those systems. These approaches focus on the effect
of priorities on the behavior of systems, and as opposed to our work, prio\-rities are not obtained
algorithmically. Moreover, priorities in~\cite{Gossler,Gossler2,Gossler3} can be considered 
stateless as they unnecessarily impose global restrictions on system actions. 
The approach in~\cite{Basu} uses priorities to control the execution of distributed systems, in order to meet 
given scheduling policies. This approach collects information wrt.\ the position of processes
at each reachable state of the system. This information is used to determine, which transitions are enabled and, 
according to the underlying scheduling policy, which transition should be executed first.
Priorities as used in~\cite{Basu} can be considered stateless, and for this reason that approach
requires to check from each reachable state, which transition should be executed first. 
Our approach also collects information from states  
that helps us to determine which transition should be executed first. However,
we collect it from preError states, and this reduces significantly the number of states that we check.
Thus, preErrors contribute to the efficiency of our approach.

The approaches in~\cite{Sifakis,Gossler} introduce dedicated components, i.e.\ \emph{schedulers},
for implementing priority mechanisms. Schedulers often introduce a number of new executions to the underlying system
which induce new reachable states. These new states, in the worst-case, multiply the size of the state space 
of the system. Our approach avoids increasing the number of reachable states of a system, 
by implementing stateful priorities directly in existing components of that system. 
Although states of transformed systems are bigger, because we introduce a number of positional variables 
(linear in the number of system components), this does not introduce new reachable states. 
%
%Algorithms for synthesizing priorities in component-based systems are proposed in~\cite{Cheng,Cheng2}.
Synthesis of priorities is reduced in~\cite{Cheng} to an \emph{EFSMT} pro\-blem, where prio\-rities 
are determined by witnesses as constructed by an EFSMT solver. 
The encoding of component-based systems proposed in~\cite{Cheng} res\-tricts the use of data variables.
The approach in~\cite{Cheng2} encodes a component-based system and an error specification as a logical formula. 
This formula is used for collecting states induced by actions that unavoidably lead to the error. 
A next step collects reacha\-ble states induced by actions alternative to those leading to the error. 
Stateless priorities are then obtained from these two types of actions. 
In this approach, the use of data variables is allowed, but only of the boolean type, and with the restriction that 
there is no data transfer in component interactions. 
We consider these restrictions too strong, since data transfer using data variables is
a typical communication way in component-based systems. The approaches in~\cite{Cheng,Cheng2}
cannot be applied in our case studies, since each of them uses, for instance, integer data variables. 
Moreover, using priorities as obtained in~\cite{Cheng,Cheng2} unnecessarily impose global restrictions on system actions.\\

\textbf{Conclusions and Future Work}. We introduced the notion of stateful prio\-rities 
for imposing precise restrictions on system actions in order to meet
a given constraint. Stateful priorities exclusively
restrict erroneous system behavior as specified by the constraint, whereas
safe system behavior remains unres\-tricted. 
We presented algorithms which are implemented in our tool CrEStO.
That tool automatically transforms networks in order to use synthesized stateful priorities.
We presented as well an upper bound formula for the number of 
reachable states in transformed networks. 
Moreover, we showed that our approach is correct in the sense of not introducing deadlocks, 
and making error states unreachable.
Our experiments with three real-world examples 
demonstrated the applicability of our approach.
We plan to extend the query language in order to support LTL properties.
We plan as well to extend this approach for timed systems. 
 
\textbf{Acknowledgments}. We would like to thank Harald Ruess and 
Hern\'an Ponce de Le\'on for their support 
and enriching suggestions during the deve\-lopment of this work.

\bibliographystyle{unsrt}
\bibliography{main}

\end{document}